\documentclass[aps,prl,twocolumn,superscriptaddress,showpacs,floatfix]{revtex4}

\usepackage{bm}
\usepackage{amsmath}
\usepackage{amssymb}
\usepackage{graphicx}
\usepackage{color}
\usepackage{xcolor}\definecolor{ao}{rgb}{0.0,0.0,1.0}

\definecolor{br}{rgb}{1.0, 0.22, 0.0}

\def\sig{{\mbox{\boldmath{$\sigma$}}}}
\input{epsf}
\begin{document}
\title{Rashba splitting of Cooper pairs\\
}

\author{R. I. Shekhter}
\affiliation{Department of Physics, University of Gothenburg, SE-412
96 G{\" o}teborg, Sweden}

\author{O. Entin-Wohlman}
\affiliation{Raymond and Beverly Sackler School of Physics and Astronomy, Tel Aviv University, Tel Aviv 69978, Israel}
\affiliation{Physics Department, Ben Gurion University, Beer Sheva 84105, Israel}

\email{oraentin@bgu.ac.il}

\author{M. Jonson}
\affiliation{Department of Physics, University of Gothenburg, SE-412
96 G{\" o}teborg, Sweden}
\affiliation{SUPA, Institute of Photonics
and Quantum Sciences, Heriot-Watt University, Edinburgh, EH14 4AS,
Scotland, UK}

\author{A. Aharony}
\affiliation{Raymond and Beverly Sackler School of Physics and Astronomy, Tel Aviv University, Tel Aviv 69978, Israel}
\affiliation{Physics Department, Ben Gurion University, Beer Sheva 84105, Israel}

\date{\today}

\begin{abstract}
We investigate theoretically the properties of a weak link between two superconducting leads, which has the form of a non-superconducting nanowire with a strong Rashba spin-orbit coupling caused by an electric field. In the Coulomb blockade regime of single-electron tunneling, we find that
such a weak link acts as a ``spin splitter" of the spin states of Cooper pairs tunneling through the link, to an extent that depends on the direction of the electric field.
We show that the Josephson current  is sensitive to interference between the resulting two transmission channels, one where the spins of both members of a Cooper pair are preserved and one where they are both flipped.
As a result, the current is a periodic function of the strength of the spin-orbit interaction and of  the  bending angle of the nanowire (when mechanically bent); an identical effect appears due to strain-induced  spin-orbit coupling. In contrast, no spin-orbit induced interference effect can influence the current through a single weak link connecting two normal metals.
\end{abstract}


\maketitle

Superconducting spintronics \cite{Linder.2015}
is of great
interest in modern research on nanophysics, due to its potential for offering both spin control
and dissipationless spin transport.
In this Letter we consider the Josephson current between two  superconductors  connected by a  weak link.  The  link suppresses the pairing potential,
thus allowing
external manipulation of the spin structure of the Cooper pairs. An obvious way to achieve this  is to use the Rashba spin-orbit (SO) coupling \cite{Rashba.1960,Winkler.2003}, which can be generated by an external electric field. The SO coupling  rotates the electron spin around an axis fixed by the electron momentum and the electric field \cite{Datta.1990}.
Quite a number of papers investigated the effect of this interaction on the  Josephson current
\cite{Linder.2002, Krive.2004.2005,
Martin.2007.2009,
Reynoso.2008, Buzdin.2008.2015,Jacobsen, Campagnano.2015,Beri.2014,Dimitrova.2006,Samokhvalov.2014}.
Most found that modifying the Josephson current by the SO interaction necessitates  breaking  time-reversal symmetry,  e.g.,  by  a magnetic-field induced Zeeman splitting
\cite{Linder.2002, Krive.2004.2005, Martin.2007.2009, Reynoso.2008, Buzdin.2008.2015, Campagnano.2015} or by magnetic exchange interactions \cite{Buzdin.2008.2015,Jacobsen,Samokhvalov.2014}.
Here we achieve the desired spin control {\it without any magnetic field} \cite{Reeg.2015}.

A mechanically bent one-dimensional (1D) nanowire subjected to a strong Rashba SO interaction and suspended between two normal bulk conductors was considered in Ref.~\onlinecite{Shekhter.2013.2014}, where it was shown that none of the three components of the electronic spin is a good quantum number. Hence, the transfer of electrons through
this {\em non-superconducting} weak link results in a split of the spinor wave-function with respect to the two possible spin projections of the incident electrons. Since the dynamic transformation of the spin caused by such a ``spin-splitter"  is fully deterministic
-- in contrast to the stochastic spin flips due to  magnetic impurities --  one may expect consequences in the form of various interference phenomena.
Here we show that the  splitting of the spin state of the paired electrons that carry the Josephson current
may transform the spin-singlet Cooper pairs into a coherent mixture of singlet and triplet spin states.
This mixture gives rise to  interference  between the channel in which both electrons preserve their spins and the channel where they are both
flipped. The resulting  interference  pattern, that appears in  the Josephson current but does not show up in the normal-state transmission of the junction, allows for electrical and mechanical control of the Josephson current between spin-singlet superconductors,  corresponding to
a new type of ``spin-gating" \cite{Shekhter.2016} of superconducting ``weak links".
\begin{figure}
\vspace{0.cm} \centerline {\includegraphics[width=11cm]{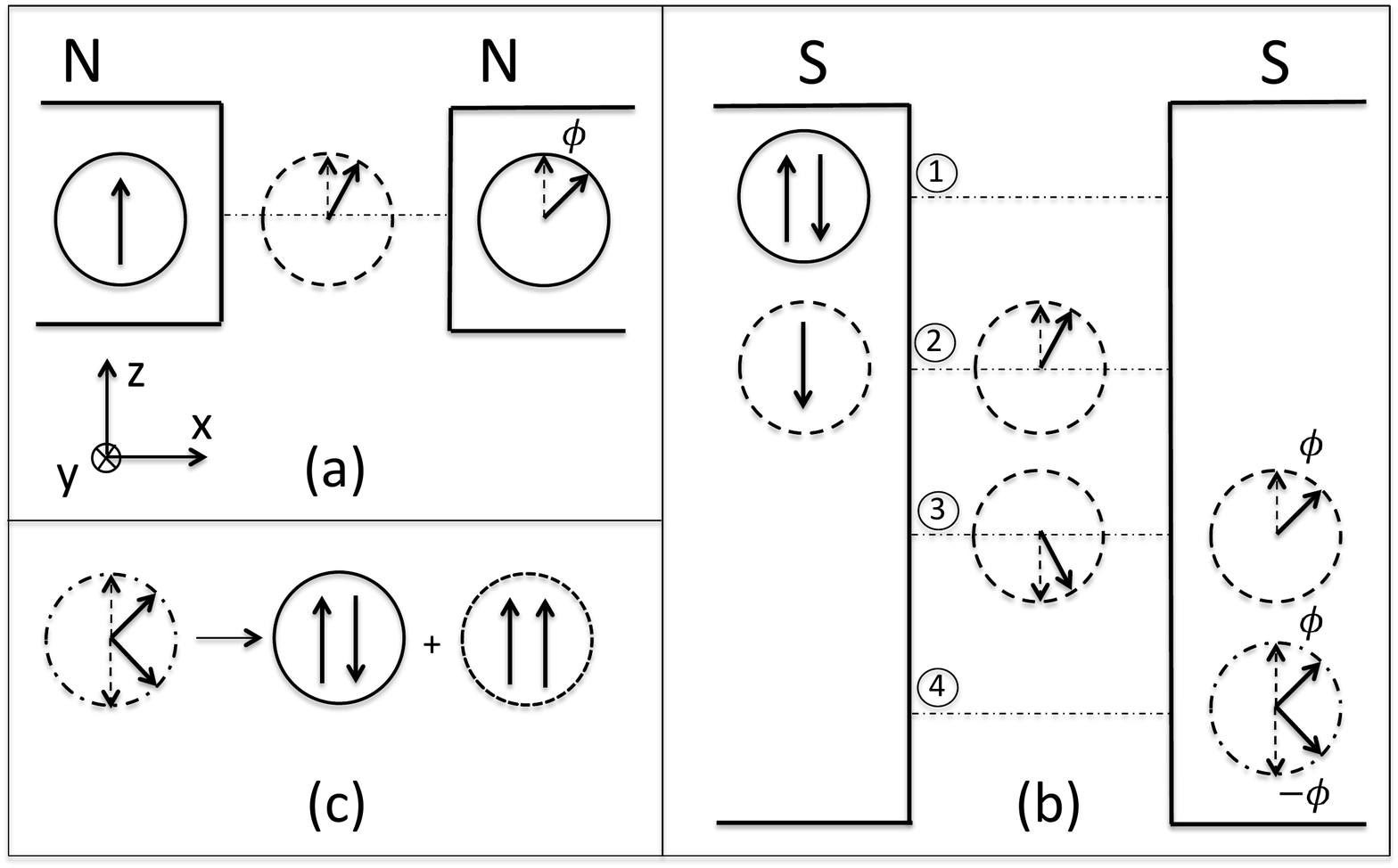}}
\vspace*{-1.5 cm}
\caption{Schematic illustrations of the lowest-order perturbation expansion steps for tunneling (in the Coulomb blockade regime) through a straight nanowire weak link 
subjected to the Rashba spin-orbit (SO) interaction
caused by an electric field along the $\hat{\bf z}-$direction. In a semiclassical picture, the spin of each electron (
arrows) is rotated in the $XZ-$plane 
as it goes through the link.
(a)
Single electron tunneling from one normal metal to another, via an intermediate (rotating) state (dashed circle).  When the electron enters the second normal metal, its spin has been rotated. 
(b) Sequential tunneling in four steps of a Cooper pair between two superconductors connected by the same weak link.
Because the two electrons that form the Cooper pair are in time-reversed states,
the SO interaction rotates their spins in opposite directions. (c) As they enter the second superconductor, the Cooper pairs are in a coherent mixture (dash-dotted circle) of a spin-singlet and a spin-triplet state. 
Inside this superconductor, this state is then
projected onto the singlet state (full circle). }
\label{Fig1}
\end{figure}

To illustrate our calculation, Fig.~\ref{Fig1}
uses a semiclassical analogue of the quantum
evolution of the spin states of electrons which move between two bulk leads via a weak link, where they are subjected to the Rashba SO interaction. For simplicity we assume for now that the weak link is a straight 1D wire along the $\hat{\bf x}-$axis. The  SO interaction in the wire is due to an  electric field, which for the moment is assumed to point along $\hat{\bf z}$ and therefore corresponds to an effective SO-interaction-induced magnetic field  directed along  $\hat{\bf y}$.
Figure~\ref{Fig1}(a) illustrates  a single-electron  transfer from one normal metal to another.  Without loss of generality, we choose the $\hat{\bf z}-$axis  in spin space to be along the direction of the polarization of the electron in the first (left) metal.
Semiclassically, the spin of the injected electron will rotate in the $XZ-$plane
as it passes through the wire.
As a result, the spins of the electrons that enter the second metal from the wire are rotated around the $\hat{\bf y}-$axis by an angle
proportional to the strength of the SO interaction and the length of the wire. This rotation depends on the direction of the ``initial" electron's polarization. It occurs only if the polarization has a component in the $XZ-$plane.
Quantum mechanically, the electron's spinor in the left metal is an eigenfunction of the Pauli spin-matrix $\sigma_z$, and the spinor of the outgoing electron is in a coherent superposition of spin-up and spin-down eigenstates of $\sigma_z$.

How can this picture be generalized to describe the transfer of the two electrons of a Cooper pair between two bulk superconductors?
The simplest case to consider, which we focus upon below, is when single-electron tunneling is Coulomb-blockaded throughout the wire. While the blockade can be lifted for one electron, double electron occupancy of the wire is suppressed, i.e.,
a Cooper pair is mainly transferred sequentially,
as shown in Fig.~\ref{Fig1}(b).
Each electron transfer is now accompanied by the  spin rotation shown in Fig.~\ref{Fig1}(a). However, since the two transferred electrons
are in time-reversed quantum states, the time evolution of their spins are reversed with respect to one another, and their rotation angles have opposite signs [step 4 in Fig. \ref{Fig1}(b)].
This final state [Fig. \ref{Fig1}(c)] can be expressed as a coherent mixture of a spin-singlet and a spin-triplet state, but only the former can enter into the second superconductor. As we show below, this projection onto the singlet causes a reduction of the Josephson current.

We consider a model where a Cooper pair is transferred
between superconducting source and drain leads via virtual states localized in a weak-link wire   [see Fig.~\ref{Fig2}(a)].
 The corresponding tunneling process, which
 supports multiple tunneling channels, was analyzed in detail in Ref.~\onlinecite{Gisselfalt.1996}. A significant simplification occurs in the Coulomb-blockade regime, defined by the inequality{
$E_{e}=E_{C}(N+1)-E_{C}(N)\gg |\Delta |$,
where $|\Delta|$  is the  energy gap parameter in the superconducting leads \cite{note}, and $E_C(N)$  is the Coulomb energy of the wire when it contains $N$ electrons. In this regime tunneling channels requiring two electrons to be simultaneously localized in a virtual state in the wire can be neglected, hence the sequential transmission. Another simplification follows from our assumption that the length of the wire $d$ is short compared to the superconducting coherence length $\xi_0 \equiv \hbar v_{\rm F}/|\Delta|$ \cite{note}, so that the dependence of the matrix element for a single electron transfer on the electron energy in the virtual states can be ignored. More details are given in Ref.~\onlinecite{SM}.
A final simplification, facilitated by the device geometry,   concerns the conservation of the electrons' longitudinal momenta as they tunnel between the two leads.
In Fig. \ref{Fig2}, the  wire ends are placed on top of the metal leads and separated from them by thin but long tunneling barriers. Since the direction of tunneling is nearly perpendicular to the direction of the current along the wire, such a geometry is conducive to longitudinal momentum conservation
\cite{Eaves.2013}.

\begin{figure}
\vspace{0.cm} \centerline {\includegraphics[width=7cm]{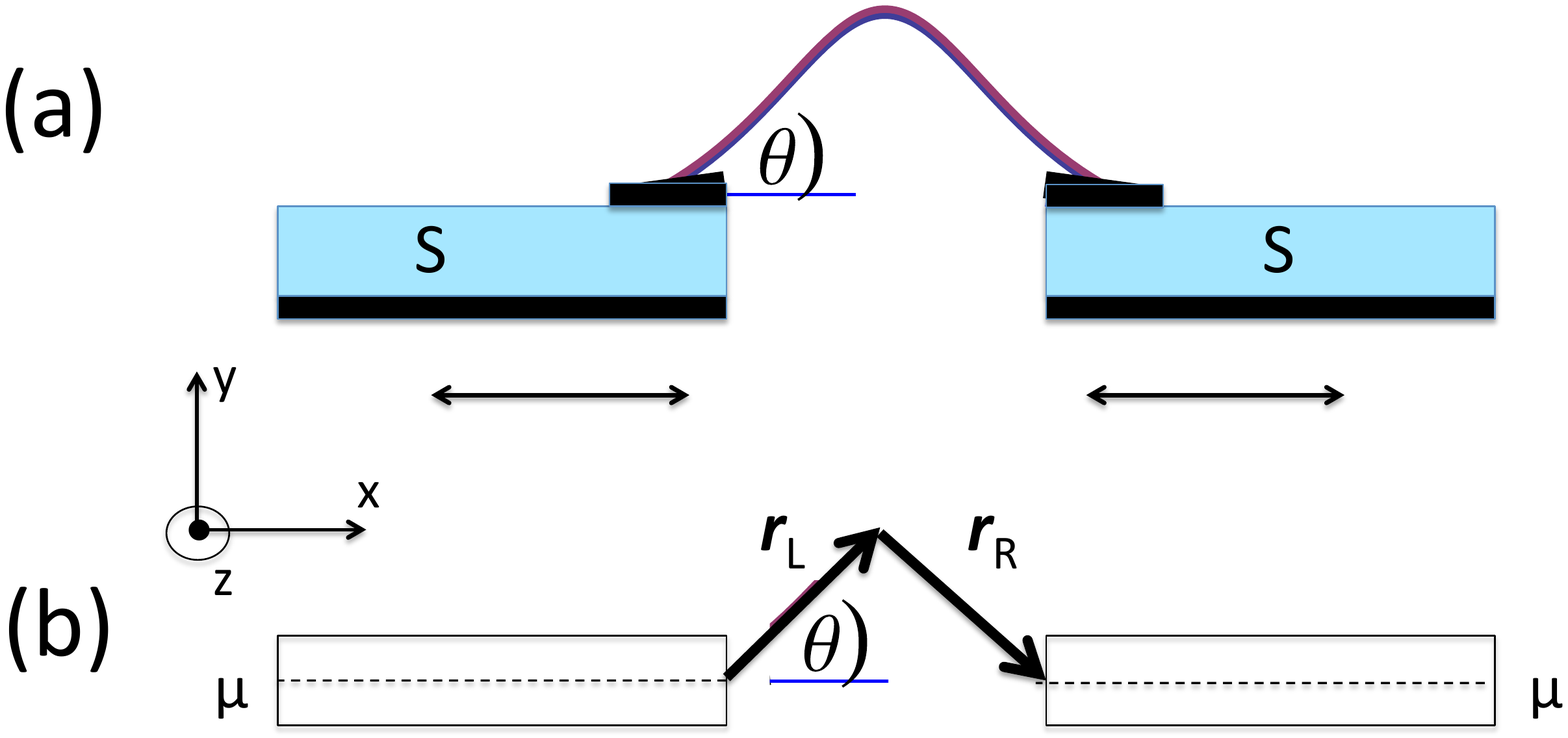}}
\vspace*{-1.5 cm}
\caption{(color online)
Sketch (a) and simplified model (b) of a
device that would allow the effects predicted in the text  to be studied.
}
\label{Fig2}
\end{figure}

These simplifying but realistic assumptions  allow us to describe the transfer of a Cooper pair between the two superconductors  in terms of
single-electron tunneling,
 as  given by the Hamiltonian ${\cal H} = {\cal H}^{}_L + {\cal H}^{}_R + {\cal H}^{}_T$.
Here
\begin{align}
\label{eq:4}
{\cal H}^{}_{L(R)} &= \sum_{{\bf k}({\bf p})} \xi^{}_{{\bf k}({\bf p})} c^\dag_{{\bf k}({\bf p})\sigma}c^{}_{{\bf k}({\bf p})\sigma} \nonumber \\
&+\Big ( \Delta_{L(R)} \sum_{{\bf k}({\bf p})} c^\dag_{{\bf k}({\bf p})\uparrow}c^\dag_{{-\bf k}(-{\bf p}),\downarrow} + {\rm H.c.} \Big )\ ,
\end{align}
where
$\xi_{{\bf k}({\bf p})} = \epsilon_{{\bf k}({\bf p})} - \mu$,
is the quasi-particle energy in the left (right) bulk superconducting lead with order parameter $\Delta_{L(R)}$, and $\mu$ is the common chemical potential.
Single-electron tunneling between the leads is described by the tunnel Hamiltonian
\begin{align}
\label{eq:5}
{\cal H}^{}_{T}=\sum_{{\bf k}, {\bf p}} \sum_{\sigma, \sigma'}
\big ( c^\dag_{{\bf p}\sigma'} [ W^{}_{{\bf p},{\bf k}} ]_{\sigma', \sigma}^{} c^{}_{{\bf k}\sigma} + {\rm H.c.} \big ) \ ,
\end{align}
where  
$[ W_{{\bf p},{\bf k}} ]_{\sigma , \sigma'} = \big ([ W_{- {\bf p}, -{\bf k}}]_{-\sigma,-\sigma'}\big )^{\ast} $ are elements of a matrix in spin space, which
obey time-reversal symmetry
\cite{Anderson.1964}.
The operator $c^\dag_{{\bf k}({\bf p})\sigma}$ creates an electron in the left (right) lead, with momentum ${\bf k}({\bf p})$ and a spin index $\sigma$, which denotes the eigenvalue of the spin projection along an arbitrary axis.
The Cooper pairs in both superconducting leads are formed by electrons whose spin projections on their common quantization axis have opposite signs,  denoted by $\sigma$ and $\bar\sigma$.
The Hamiltonian (\ref{eq:5}) describes how single electrons are transferred through the nanowire weak link
while their spin states evolve under the  SO interaction, which is geometrically confined to the wire.
Such tunneling amplitudes, for a normal-state junction, were considered before \cite{Raikh.1994,Ora.and.Amnon}.

We assume a weak link containing a bent wire [see Fig.~\ref{Fig2}(a)]. The actual calculations are done for the geometry shown in Fig.~\ref{Fig2}(b), where the weak link comprises two straight 1D wires, ${\bf r}_L$ and ${\bf r}_R$, of equal length $d/2$,  connected by a ``bend". The angles between these wires and the $\hat{\bf x}-$axis are $\theta$ and $-\theta$, respectively. As
shown below, the resulting Josephson current depends on
$\theta$, allowing its mechanical manipulation.  The Rashba Hamiltonian for a wire lying in the $XY-$plane is
\begin{align}
{\cal H}^{}_{\rm wire}=-\frac{1}{2m^*}\frac{d^{2}}{d{\bf r}^{2}}-i\frac{k^{}_{\rm so}}{m^*}\hat{\bf n}\cdot\sig\times
\frac{d}{d{\bf r}}\ ,
\label{HAM}
\end{align}
where ${\bf r}$ is the coordinate  along
a straight wire segment,  $\sig$ is the vector of the Pauli matrices, $m^*$ is the effective electronic mass, and $k_{\rm so}$  is the strength of the  SO interaction
in momentum units (using $\hbar=1$), related to an electric field along  $\hat{\bf n}$ that generates the SO  coupling \cite{Rashba.1960,Winkler.2003,Datta.1990}.
For this configuration,
the tunneling amplitude, a $2\times 2$ matrix in spin space, is \cite{SM}
\begin{align}
W^{}_{{\bf k},{\bf p}}
/W^{}_{0}\equiv{\cal W}, \quad
{\cal W} = e^{-i{k}_{\rm so}\sig\cdot {\bf r}^{}_{L}\times\hat{\bf n}}
e^{-i{k}_{\rm so}\sig\cdot {\bf r}^{}_{R}\times\hat{\bf n}}.
\label{WLR}
\end{align}
In Eq.~(\ref{WLR}),
$W_{0}= G_{0}( |{\bf r}_{L}|;E)
{\cal T}
G_{0}( |{\bf r}_{R}|;E)
$
comprises all the characteristics of the tunneling matrix that are independent of the spin dynamics,
where
${\cal T}$ is
the transfer matrix through the bend in the wire,
and
$G_{0}( |{\bf r}|;E)=-i(\pi m/\widetilde{\kappa})\exp[i\widetilde{\kappa} |{\bf r}|]$
is the free propagator modified by the SO interaction according to
$\widetilde{\kappa}=(2mE+\widetilde{k}_{\rm so}^{2})^{1/2}$. Here
$\widetilde{k}_{\rm so}=k_{\rm so}\sqrt{1-(\hat{\bf n}\cdot\hat{\bf u})^{2}}$,
where
$\hat{\bf u}$ is a unit vector along the direction of the straight segment.
The unitary matrix ${\cal W}$
performs two consecutive spin rotations 
around the 
axes ${\bf r}_{L}\times\hat{\bf n}$ and ${\bf r}_{R}\times\hat{\bf n}$, as described (for a single segment) semiclassically in Fig.~\ref{Fig1}(a).
Equation 
(\ref{WLR}) 
is derived to lowest possible order in the tunneling; the explicit dependence of $W_{0}$ 
on the momenta
are omitted for brevity \cite{SM}.

The SO interaction modifies significantly the amplitude of the Josephson equilibrium current, while leaving the transmission of the junction in its normal state as
in the absence of this coupling. The matrix 
${\cal W}$, that determines these quantities,  
depends crucially on the direction $\hat{\bf n}$ of the electric field.
In the ``genuine" Rashba configuration,  $\hat{\bf n}$ is normal to the plane of the junction; this is the case described semiclassically in Fig.~\ref{Fig1}.
For $\hat{\bf n}\parallel\hat{\bf z}$, 
\begin{align}
&{\cal W}
=
\big[\cos^{2}(k^{}_{\rm so}d/2)-
\sin^{2}(k^{}_{\rm so}d/2)
\cos (2\theta)\big]\nonumber\\
&+i\sig\cdot\big [\hat{\bf y}
\sin(k^{}_{\rm so}d)\cos(\theta)
+\hat{\bf z}\sin^{2}(k^{}_{\rm so}d/2)\sin (2\theta)
\big ]\ .
\label{WLR1}
\end{align}
In contrast, when the electric field is in the plane of the junction, e.g.,
$\hat{\bf n}=\hat{\bf y}$, we find
\begin{align}
{\cal W}=
\cos [k^{}_{\rm so}d\cos(\theta)]-i\sig\cdot\hat{\bf z}\sin [k^{}_{\rm so}d\cos(\theta)]\ .
\label{WLR2}
\end{align}
Spin-orbit coupling may also be induced by {\em strain} \cite{referee};
in that case the last term in Eq.  (\ref{HAM}) is replaced by  ${\cal H}_{\rm so}=(\Delta^{\rm strain}_{\rm so}/2)\hat{\bf k}\cdot\sig$, where $\hat{\bf k}$ is a unit vector along the
momentum (i.e., along the straight segment) \cite{Rudner.Rashba.2010,SOcomment}. By defining
$\Delta
^{\rm strain}_{\rm so}/2=\hbar v_{\rm F}k^{\rm strain}_{\rm so}$,
 one finds that ${\cal W}$ of the strain-induced case has the same form as
Eq. (\ref{WLR1}), except that $\hat{\bf y}$  is replaced by $\hat{\bf x}$. The resulting expressions for  the Josephson current and for the  normal-state transmission turn out to be the same as  for the
SO interaction in the
 ``genuine" Rashba configuration.

A calculation of the current emerging from the left lead, 
$I_{L}=-e\langle \dot{N}_{L}\rangle =-ie
 \langle [ {\cal H}, \sum_{{\bf k}, \sigma} c^{\dagger}_{{\bf k} \sigma}c^{}_{{\bf k} \sigma} ]\rangle
$,  
up to second order in ${\cal H}_{T}$, is straightforward \cite{SM}.
It differs from the standard procedure \cite{Kulik} only in that the spin dynamics caused by the SO interaction must be properly taken into account  by allowing the tunneling of electrons to be spin dependent. 
The current  
comprises the supercurrent 
$J(\varphi)$ and the quasi-particle current. 

In the absence of the SO interaction, the quasi-particle current scales as the transmission of the junction when in the normal state \cite{Ambegaokar.Baratoff,Kulik}.  The SO coupling modifies this transmission by  the factor  
Tr$\{{\cal W}{\cal W}^\dag\}$,  
where the trace is in spin space \cite{SM}.  This  factor is simply 2, the spin degeneracy [see Eq. (\ref{WLR})]; i.e., the SO interaction does not affect the electric conductance (unless the junction allows for geometrically-interfering processes \cite{Ora.and.Amnon}).
The superconducting Josephson current is
\begin{align}
\frac{J(\varphi)}{J^{}_0(\varphi)} =\frac{1}{2}
\sum_\sigma\left[
\vert
{\cal W}_{\sigma\sigma}\vert^2 -\vert
{\cal W}_{\sigma\bar{\sigma}}\vert^2 \right]
=
\sum_\sigma\left[
\frac{1}{2}-\vert
{\cal W}_{\sigma\bar{\sigma}}\vert^2
 \right]\ ,
\label{eq:13}
\end{align}
where 
$J^{}_0(\varphi)\propto \sin(\varphi)$ is the equilibrium Josephson current in the absence
of the SO interaction  \cite{Ambegaokar.Baratoff}, and $\varphi$ is the superconducting phase
difference. 

 The matrix element
${\cal W}_{\sigma\bar{\sigma}}$
depends on the quantization axis of the spins. Choosing this axis to be along $\hat{\bf z}$,
 the ``genuine" Rashba configuration implies that $\hat{\bf n}\parallel\hat{\bf z}$, and  Eq. (\ref{WLR1}) yields
({\it cf.} Fig. \ref{Fig3})
 \begin{align}
\label{eq:17}
1 -2 \vert
{\cal W}_{\sigma\bar{\sigma}
}
\vert^2=1-2\cos^{2}(\theta)\sin^{2}(k^{}_{\rm so}d)\ .
\end{align}
The Josephson current 
is thus significantly modified.
In contrast,
 when the electric field is in the plane of the junction,
$\hat{\bf n}=\hat{\bf y}$, the matrix 
${\cal W}$
is diagonal [Eq. (\ref{WLR2})],
$J(\varphi)=J_0(\varphi)$, and
the superconducting current is not affected by the spin dynamics.
Similar qualitative results are found \cite{SM} for all the directions of the spin quantization axis.
The splitting of the Cooper-pair spin state by the SO 
interaction reduces
the Josephson current through 
the superconducting weak link under consideration.

\begin{figure}
\vspace{0.cm} \centerline {\includegraphics[width=7.5cm]{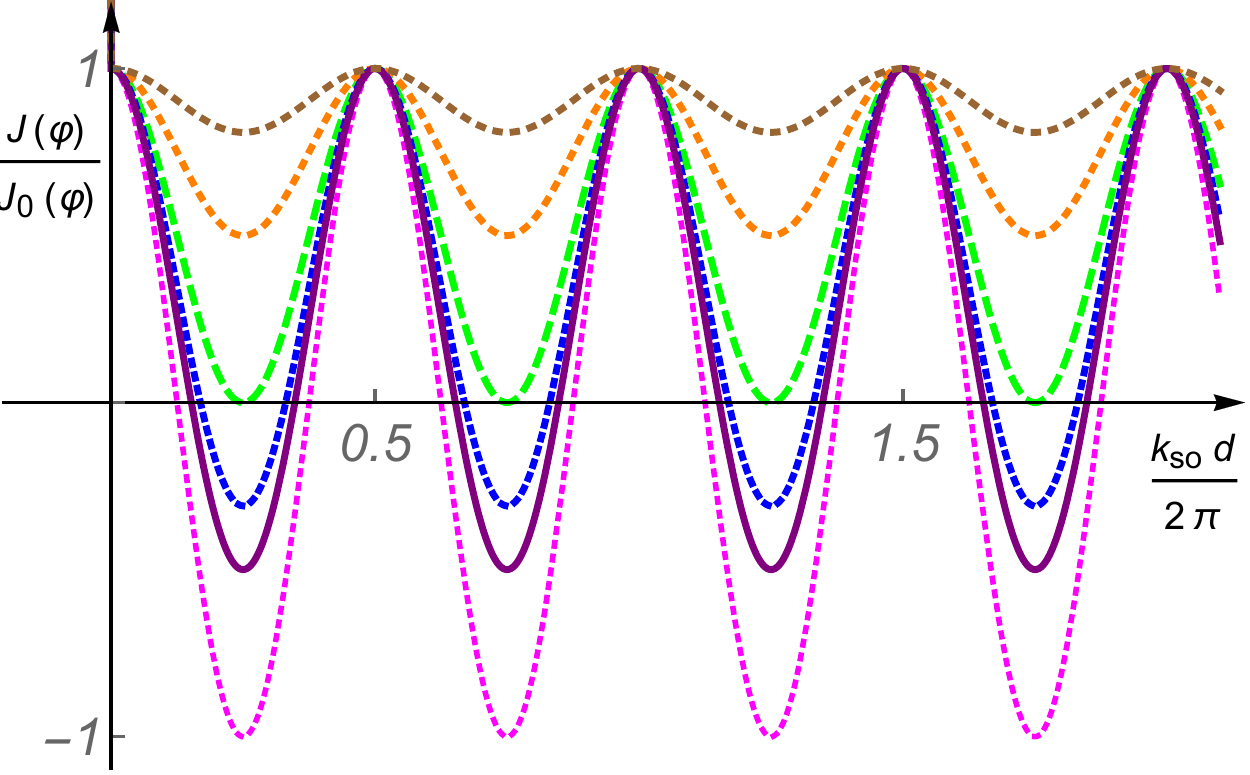}}
\caption{(color online)
The Josephson current $J(\varphi)$  divided by its value without the SO interaction, $J_{0}(\varphi)$,
for the ``genuine"  Rashba configuration
[Eqs.~(\ref{eq:13})-(\ref{eq:17})] 
as a function
of $k_{\rm so} d/(2\pi)$. The largest amplitude is  for zero bending angle, $\theta=0$, decreasing   gradually for $\theta= \pi/6 ,\pi/5 ,\pi/4 ,\pi/3 ,\pi/2.5$  [Fig.~\ref{Fig2}(b)].
Relevant values of $k_{\rm so}$ are estimated in the text.
}
\label{Fig3}
\end{figure}

Two features determine the magnitude of the effect for a given spin quantization axis in the leads (in addition to the strength $k_{\rm so}$ of the SO interaction and the length $d$ over which it acts). One is the extent to which the nanowire is bent 
[$\theta$ in Fig. \ref{Fig2}(b)],  and the other is the orientation $\hat {\bf n}$ of the electric field responsible for the SO coupling  relative to the spin quantization axis. Both 
break spin conservation, which results in Rabi oscillations between the singlet  and triplet spin states of the (originally spin-singlet) Cooper pairs passing 
through the spin-orbit-active weak link.  The consequence is a spin splitting of the Cooper pairs that reach the second superconducting lead, where their spin state is projected onto the singlet state.
This splitting can result in a Josephson current that is an oscillatory function of the ``action" $k_{\rm so} d$ of the SO interaction  (Fig. \ref{Fig3});  
the effect may be absent for special directions of the electric field.  Both results can 
be understood in terms of a semi-classical picture, 
Fig. \ref{Fig1}.

As seen in Eq. (\ref{eq:13}), the Josephson current can be written as a sum of two contributions. 
One, $\vert
{\cal W}_{\sigma\sigma}
\vert^2$, comes from a channel where the spin projections of the Cooper pair electrons, when leaving and entering 
the weak link, 
are identical; the other,
$\vert
{\cal W}_{\sigma\bar{\sigma}
}
\vert^2$, arises from  another channel, where both electron spins are flipped
during the passage. It is remarkable that the two contributions have opposite signs. This is due to  a Josephson tunneling ``$\pi$-shift"
caused by electronic spin flips 
(and is similar to the effect predicted for tunneling through a Kondo impurity \cite{Glazman.1989.Spivak.1991}). 
In particular,  a total cancellation of the Josephson current is possible when, e.g., 
$\theta = 0$ and $k_{\rm so} d = \pi/4$;  
in the limit $\theta=0$ and $k_{\rm so} d = \pi/2$ the Josephson current even changes its
sign.
This spin-orbit induced interference effect on the Josephson current is specific to a   weak link  subjected to SO interaction} between superconductors. There is no such effect on the current through a single weak link connecting two normal metals.

According to Eq.~(\ref{eq:13}), none or both of the Cooper pair electrons must have flipped their spins as they leave the weak link in order to contribute to the Josephson current. This is because 
only spin-singlet Cooper pairs can enter the receiving $s$-type bulk superconductor. However,  single-flip processes, where only one of the two tunneling electrons flips its spin, are also possible results of injecting Cooper pairs into a Rashba weak link. Those processes correspond to a triplet component of the spin state of the transferred  pair,  and  can be viewed as
evidence for  {\em spin polarization of  injected Cooper pairs}. The triplet component  could be responsible for a spin-triplet proximity effect
\cite{Reeg.2015}
and would presumably contribute a spin supercurrent 
if higher-order tunneling processes were taken into account.

In conclusion,  we have shown that 
the Josephson current through a 
weak-link  nanowire,  locally subjected to the Rashba SO 
interaction,  is sensitive to both the amount of bending and to the orientation of the electric field generating the SO coupling. This allows for a tuning of the supercurrent by mechanical and electrical manipulations of the spin polarization of the Cooper pairs. 
In particular the Josephson current through an electrostatically-gated device becomes an oscillatory function of the gate voltage.
We emphasize that these results follow from the interference of two transmission channels, one where the spins of both members of a Cooper pair are preserved and one where they are both flipped, and that
this interference does not require any external magnetic field.
It is important, however, that those parts of the device where the superconducting pairing potential is non-zero and where the SO coupling is finite are spatially separated. To lowest order in the tunneling this separation prevents the superconductivity in the leads to have any effect on the dynamical spin evolution in the wire.

Carbon nanotubes and semiconductor wires 
seem particularly suitable to be used as  spin-splitters.
Measured Rashba  spin-orbit-coupling induced energy gaps in
InGaAs/InAlAs ($\Delta_{\rm so}=2\hbar v_{\rm F}k_{\rm so}\approx 5$~meV) \cite{Nitta.1997} and 
InAs/AlSb ($\Delta_{\rm so} \approx 4$~meV) \cite{Heida.1998} quantum wells
correspond to 
$k_{\rm so} \approx 4\times 10^6$ m$^{-1}$. 
The strain-induced SO energy gap for a carbon nanotube  is
$\Delta^{\rm strain}_{\rm so}=2\hbar v_{\rm F}k^{\rm strain}_{\rm so}\approx 0.4$ meV, corresponding
to  $k^{\rm strain}_{\rm so}\approx 0.4\times 10^6$ m$^{-1}$ for $v_{\rm F}\approx 0.5\times 10^6$ m/sec \cite{McEuen.2008}.
For 
$d$ of the order of $\mu$m,
 $k^{\rm (strain)}_{\rm so}  d$ can therefore be of order $1-5$.

\begin{acknowledgments}
This work was partially supported by the Swedish Research
Council (VR), by the Israel Science Foundation
(ISF) and by the infrastructure program of Israel
Ministry of Science and Technology under contract
3-11173.
OEW and AA
thank the Department of Physics at the University of Gothenburg
for hospitality. \end{acknowledgments}


\begin{thebibliography}{99}

\bibitem{Linder.2015}
 J. Linder  and J. W. A. Robinson, 
Nat. Phys. {\bf 11}, 307 (2015).

\bibitem{Rashba.1960}
 E. I. Rashba, Fiz. Tverd. Tela (Leningrad) {\bf 2}, 1224 (1960) [Sov. Phys. Solid State {\bf 2}, 1109 (1960)]; Y. A. Bychkov and E. I. Rashba, J. Phys. C {\bf 17}, 6039 (1984).

\bibitem{Winkler.2003}
 R. Winkler, {\it Spin-Orbit Coupling Effects in Two-Dimensional Electron and Hole Systems} (Springer-Verlag, Berlin, 2003).

\bibitem{Datta.1990}
S.~Datta and B.~Das,
Appl. Phys. Lett. {\bf 56}, 665 (1990).

\bibitem{Linder.2002}
E. V. Bezuglyi,  A. S. Rozhavsky,  I.~D.~Vagner,  and P.~Wyder,
Phys. Rev. B {\bf 66}, 052508 (2002).


\bibitem{Krive.2004.2005}
I.~V.~Krive, L.~Y.~Gorelik, R.~I.~Shekhter, and M.~Jonson,
Low Temp.Phys. {\bf 30}, 395 (2004);
I.~V.~Krive, A.~M.~Kadigrobov, R.~I.~Shekhter, and M.Jonson,
Phys. Rev. B {\bf 71}, 214516 (2005).

\bibitem{Martin.2007.2009}
L. Dell'Anna,  A.~Zazunov,  R.~Egger,  and T.~Martin,
Phys. Rev. B {\bf 75}, 085305 (2007);
A.~Zazunov, R.~Egger, T.~Jonckheere and T.~Martin,
Phys. Rev. Lett. {\bf 103}, 147004 (2009);
A.~Brinetti, A.~Zazunov, A.~Kundu, and R.~Egger,
Phys. Rev. B {\bf 88}, 144515 (2013);
T.~Yokoyama, M.~Eto, Yu.~V.~Nazarov,
Phys. Rev. B {\bf 89}, 195407 (2014).


\bibitem{Reynoso.2008}
A. A. Reynoso, Gonzalo Usaj,  C.~A.~Balseiro,  D.~Feinberg,  and M.~Avignon,
Phys. Rev. Lett. {\bf 101}, 107001 (2008).

\bibitem{Buzdin.2008.2015}
A.~Buzdin,
Phys. Rev. Lett. {\bf 101}, 107005 (2008);
S.~V.~Mironov, A.~S.~Mel'nikov, A.~I.~Buzdin,
Phys. Rev. Lett. {\bf 114}, 227001 (2015).



\bibitem{Jacobsen}
S. H. Jacobsen and J. Linder,
Phys. Rev. B {\bf 92},  024501 (2015).






\bibitem{Campagnano.2015}
G.~Campagnano, P.~Lucignano, D.~Giuliano and A.~Tagliacozzo,
J. Phys. Condens. Matter  {\bf 27}, 205301 (2015).

\bibitem{Beri.2014}
B. B{\'e}ri,  J. H. Bardarson,  and C.~W.~J.~Beenakker,
Phys. Rev. B {\bf 77}, 045311 (2008).

\bibitem{Dimitrova.2006}
O. V. Dimitrova and M. V. Feigel'man,
J. Exp. Theor. Phys. {\bf 102}, 652 (2006).

\bibitem{Samokhvalov.2014}
A.~V.~Samokhvalov, R.~I.~Shekhter, and A.~I.~Buzdin,
Sci. Rep. {\bf 4}, 5671 (2014).

\bibitem{Reeg.2015}
Spin-triplet pairing correlations in a conductor with SO interactions in the proximity of an s-wave superconductor were found in the absence of a magnetic field by
C. P. Reeg and D. L. Maslov,
Phys. Rev. B {\bf 92}, 134512 (2015).

\bibitem{Shekhter.2013.2014}
R. I. Shekhter, O. Entin-Wohlman, and A.~Aharony,
Phys. Rev. Lett. {\bf 111}, 176602 (2013); R.~I.~Shekhter, O.~Entin-Wohlman and A.~Aharony,
Phys. Rev. B {\bf 90}, 045401 (2014).


\bibitem{Shekhter.2016}
R. I. Shekhter and M. Jonson,
Synth. Met. (in press); arXiv:1507.05822.

\bibitem{Gisselfalt.1996}
M. Gisself{\"a}lt,
Phys. Scr. {\bf 54}, 397 (1996); see in particular Eqs. (18) and (39).

\bibitem{note}
In order to keep the estimates simple, it is assumed that both the Fermi velocities, $v_{\rm F}$,  and the superconducting gaps of the two leads are the same.

\bibitem{SM}
See Supplemental Material at [URL will be inserted by publisher] for a detailed derivation.


\bibitem{Eaves.2013}
Since
electron tunneling
is confined to a restricted interval $\Delta x$ at one or the other end of the wire, where the overlap between the
wave functions in the lead and the wire
is finite,
there is an uncertainty $\Delta k \sim 1/\Delta x$ in the longitudinal momentum of the tunneling electrons. 
We view this uncertainty
as a splitting of the incoming well-defined momentum state into several channels $k$ within a range $\Delta k$. This enables 
co-tunneling processes,
where the momentum of the electron entering the wire from the left electrode and the momentum of the electron leaving the wire for the right electrode are different. However,
for the  geometry  in Fig. \ref{Fig2} the direction of tunneling is nearly orthogonal to that of the current.
This is why $\Delta x$  can be chosen large enough to suppress quantum fluctuations of the longitudinal electron momentum (and therefore co-tunneling). A resonant tunneling effect exploiting this 
longitudinal momentum-conservation was recently observed
by L. Britnell, R. V. Gorbachev, A.~K.~Geim, L.~A.~Ponomarenko, A.~Mishchenko, M.~T.~Greenaway, T.~M.~Fromhold, K.~S.~Novoselov, and L.~Eaves,
Nat. Commun. {\bf 4}, 1794 (2013).

\bibitem{Anderson.1964}
P. W. Anderson,
in {\em Lectures on the many-body problem}, vol. 2 p. 113, Ed. E.~R.~Caianiello (Academic Press, New York, 1964).

\bibitem{Raikh.1994}
T. V. Shahbazyan and M. E. Raikh,
Phys. Rev. Lett. {\bf 73}, 1408 (1994).

\bibitem{Ora.and.Amnon}
O. Entin-Wohlman, A. Aharony, Y. Galperin, V. Kozub, and V. Vinokur,
Phys. Rev. Lett. {\bf 95}, 086603 (2005);
A.~Aharony, Y.~Tokura, G. Z. Cohen, O. Entin-Wohlman, and S. Katsumoto,
Phys. Rev. B {\bf 84}, 035323 (2011).




\bibitem{referee}
We thank one of the referees for pointing this out.


\bibitem{Rudner.Rashba.2010}

M. S. Rudner and E. I. Rashba,  Phys. Rev. B {\bf 81}, 125426 (2010);
K. Flensberg and C. M. Marcus, Phys. Rev. B {\bf 81}, 195418 (2010).


\bibitem{SOcomment}
In this case the electric field is  a host crystal field, see
L.~Chico, M.~P.~L\'{o}pez-Sancho, and M.~C.~Mu\~{n}oz,
Phys. Rev. B {\bf 79}, 235423  (2009).




\bibitem{Kulik}
I. O. Kulik and I. K. Janson,
{\em Josephson effect in superconductive tunneling structures} (Israel Program for Scientific Translations
[Keter press], Jerusalem, 1972).

\bibitem{Ambegaokar.Baratoff}
V. Ambegaokar and A. Baratoff,
Phys. Rev. Lett. {\bf 10}, 486 (1963); erratum,
Phys. Rev. Lett. {\bf 11}, 104 (1963).

\bibitem{Glazman.1989.Spivak.1991}
L. I. Glazman and K.~A.~Matveev,
Pis'ma Zh. Eksp. Teor. Fiz.
{\bf 49}, 570 (1989) [JETP Lett. {\bf 49}, 660 (1989);
B.~I.~Spivak and S.~A.~Kivelson,
Phys. Rev. B {\bf 43}, 3740 (1991).



\bibitem{Nitta.1997}
J.~Nitta, T.~Akazaki, H.~Takayanagi, and T.~Enoki
Phys. Rev. Lett. {\bf 78}, 1335 (1997).


\bibitem{Heida.1998}
J. P. Heida, B.~J.~van Wees, J.~J.~Kuipers, T.~M.~Klapwijk, and G.~Borghs,

\bibitem{McEuen.2008}
F.~Kuemmeth, S.~Ilani, D.~C.~Ralph, and P.~L.~McEuen,
Nature {\bf 452}, 448 (2008).








\end{thebibliography}
\end{document}